\documentclass{mem}
\usepackage{natbib}
\usepackage{txfonts}
\usepackage{graphicx}
\usepackage[a4paper]{hyperref}
\usepackage{color}

\idline{1}{1}

\begin{document}

\title{Global Architecture of Planetary Systems (GAPS), a project for the whole
Italian Community}

   \subtitle{}

\author{E.\,Poretti\inst{1}, 
C.\,Boccato\inst{2}, 
R.\,Claudi\inst{2}, 
R.\,Cosentino\inst{3},
E.\,Covino\inst{4},
S.\,Desidera\inst{2},
R.\,Gratton\inst{2},
A.F.\,Lanza\inst{5},
A.\,Maggio\inst{6},
G.\,Micela\inst{6},
E.\,Molinari\inst{3},
I.\,Pagano\inst{5},
G.\,Piotto\inst{7},
R.\,Smareglia\inst{8},
A.\,Sozzetti\inst{9},
and the whole  GAPS collaboration
}
\institute{
INAF -- Osservatorio Astronomico di Brera, Via E. Bianchi 46, I-23807 Merate (LC)
\and
INAF -- Osservatorio Astronomico di Padova, Vicolo dell'Osservatorio 5, I-35122 Padova
\and
Fundaci\'on Galileo Galilei -- INAF, Rambla Jos\'e Ana Fernandez P\'erez 7, E-38712 Bre\~na Baja, TF
\and
INAF -- Osservatorio Astronomico di Capodimonte, Salita Moiariello 16, I-80131 Napoli
\and
INAF -- Osservatorio Astrofisico di Catania, Via S. Sofia 78, I-95123 Catania
\and
INAF -- Osservatorio Astronomico di Palermo, Piazza del Parlamento 1, I-90134 Palermo
\and 
Dip. di Fisica e Astronomia Galileo Galilei -- Universit\`a di Padova, Vicolo dell'Osservatorio 2, I-35122 Padova
\and
INAF -- Osservatorio Astronomico di Trieste, Via Tiepolo 11, I-34143 Trieste
\and
INAF -- Osservatorio Astrofisico di Torino, Via Osservatorio 20, I-10025 Pino Torinese
}

\authorrunning{Poretti et al.}

\titlerunning{The GAPS project}

\abstract{The GAPS project is running since 2012 with the goal to optimize the science return
of the HARPS-N instrument mounted at Telescopio Nazionale Galileo. A large number of astronomers
is working together to allow the Italian community to gain an international  position adequate to the HARPS-N 
capabilities in the exoplanetary researches. 
Relevant scientific results are being obtained on both the main guidelines of the collaboration,
i.e., the discovery surveys and the characterization studies. The planetary system
discovered around the southern component of the binary XO-2 and its characterization together
with that of the system orbiting the northern component are a good example of the completeness 
of the topics matched by the GAPS project. 
The dynamics of some planetary systems are investigated by studying the Rossiter-McLaughlin effect,
while host stars are characterized by means of asteroseismology and star-planet
interaction.

}
		\maketitle{}

\section{Introduction}
The  ``Global Architecture of Planetary Systems" (GAPS) project	
was set after a	call to	the Italian scientists interested in the field	
of exoplanetary science to share their specific knowledge.
As a result, 
the GAPS team joins now experts in high-‐resolution spectroscopy, stellar activity
and pulsations,	crowded	stellar	environments, planetary	systems	formation, planetary	
dynamics, and data handling. The team currently includes 58 astronomers from twelve
INAF structures and Italian universities. Scientific and technical issues are
discussed within a broad Science Team who gives advise to the Project Board. 
R.~Claudi (2012-13), I.~Pagano (2013-14), and G.~Micela (2014-15) were the previous
Chairpersons, while A.F.~Lanza is on duty since May 2015.

The long-term, infrastructural goal of the GAPS project is to optimize the science return of the HARPS-‐N	
open time, functional at gaining a prominent position of the Italian community in	
the international context of the exoplanetary science.	
The scientific goal of GAPS is to understand the architectures of planetary systems	
and their properties. The adopted approach  includes the 
search for new planets ({\it discovery surveys}) around stars with well-‐defined
characteristics	and the  investigation of  the diversity of orbital and physical
properties of
known planetary	systems ({\it characterization studies}), thus exploiting at best the 
outstanding capabilities of HARPS-‐N.	
The two	approaches are complementary and both relevant to tackle 
the question of the architecture of the planetary systems
in a comprehensive way. 
  
\section{Observations}
 
\begin{figure}[t!]
\resizebox{\hsize}{!}{\includegraphics[clip=true]{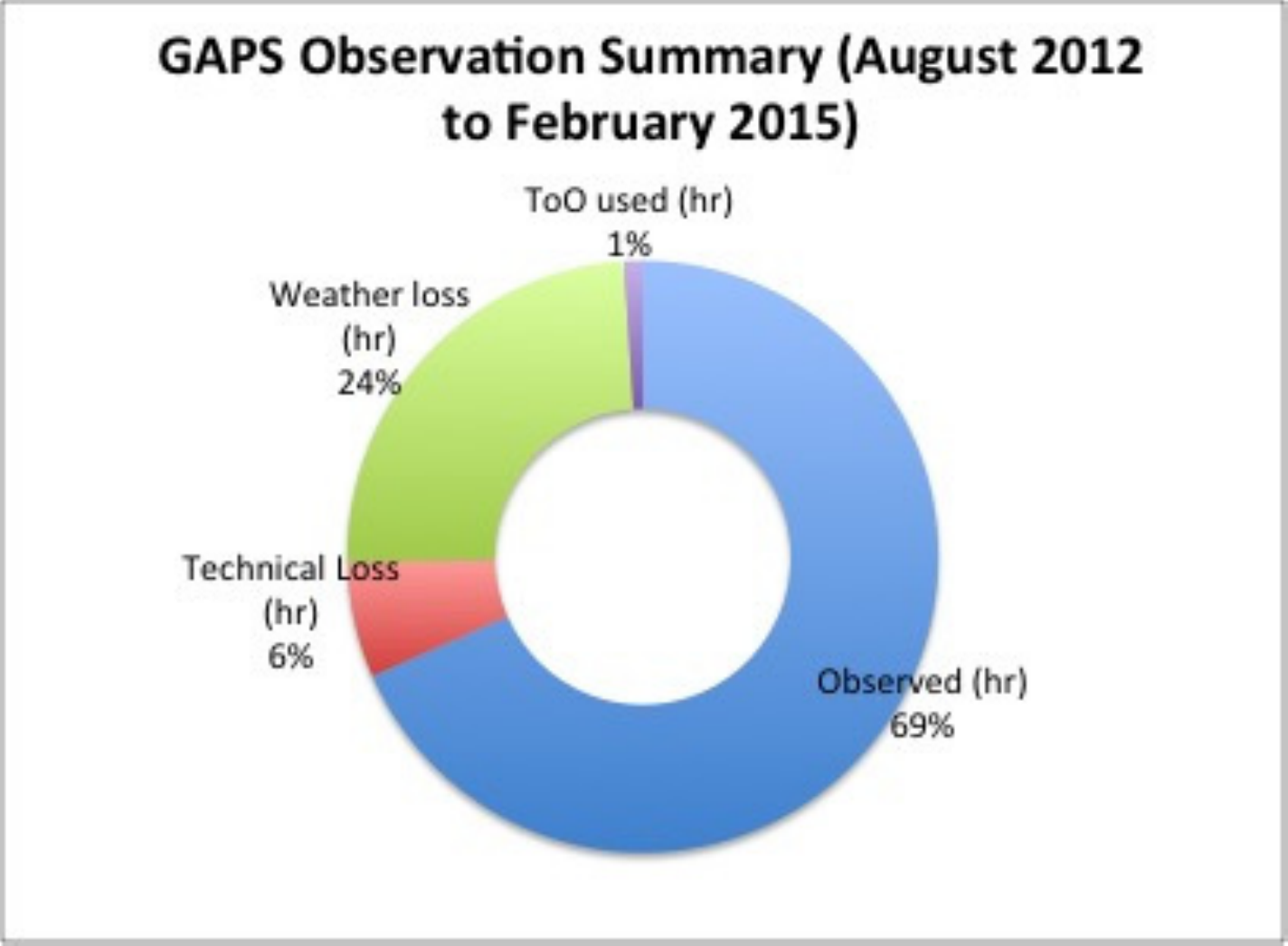}}
\caption{\footnotesize
The general summary of the use of GAPS observing time from AOT26 to AOT30.}
\label{ric}
\end{figure}

Since August 2012 (AOT26)	
236 nights have been assigned to the GAPS program (P.I. A.~Sozzetti). 
They were subdivided among the different sub-programs 
on the basis of	priority reasons. Up to now (January 31st, 2015) we collected a total of 4755	
spectra	of 273	different objects. Figure~\ref{ric} shows the statistics	
of the observing nights.
During	AOT26,	in October and November 2012, the red-side chip of HARPS-N
had a failure and the detector was substituted. Due to this, there was a
large amount of	lost nights for technical reasons (64~hrs out of 141.0~hrs).
Moreover, the instrument worked with only half chip in the remaining clear hours of 
the period (41~hrs). Just after	the HARPS-N commissioning, some technical
troubles also occurred, translating  into relevant losses of observing time. 
After the solution of several minor problems, 
the losses due to technical reasons settled down to usual values
($\sim$2\%) in the subsequent semesters.

The achieved radial-velocity precision depends on various factors, not the least the 
stellar intrinsic variability, due to pulsations and/or activity. The stellar signal
is currently indicated  as stellar noise or jitter, but often it contains relevant
physical information, like the large separation of the solar-like oscillations or
the star's rotational period. Nevertheless, even for a ``quiet" star, the uncertainties
on the radial-velocity value strongly depends on the number of spectral lines, as well as 
on their widths and depths.  In order to provide a quantitative measurement of the  radial-velocity precision,
the quiet, non-rotating star HD166620, was observed for 21 months, from June, 2012 to March,
2014 in the framework of the HARPS-N consortium.  
In this case, the radial velocity dispersion was of 1.29~m\,s$^{-1}$,
calculated on the basis of 309 spectra taken on 21 months \citep{rosario}. This value is 
in agreement with the expected value and puts in evidence the high stability of HARPS-N
over a long time baseline. 

\section{Scientific results}

\begin{figure*}[!t]
  \includegraphics[width=\linewidth]{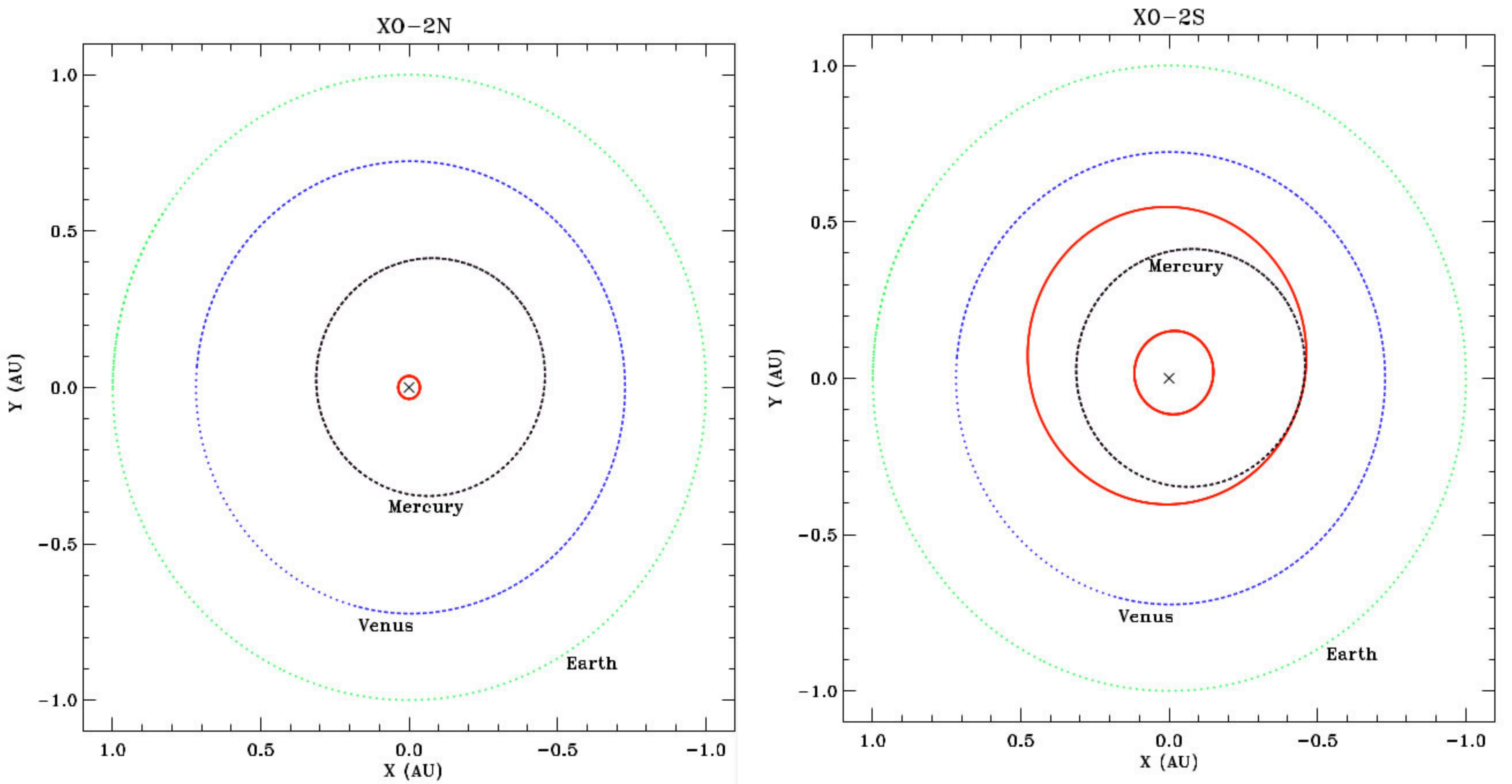}
\caption{\footnotesize Comparison of the architectures of the XO-2N (left) and XO-2S (right)
planetary systems with that of the inner Solar System. Red ellipses represent the orbits
of the XO-2 planets. }
\label{xo2ns}
\end{figure*}


An important scientific return was expected from the large amount of observing time allocated at 
the GAPS project. This was  not an obvious task when considering the high-level of competition
in the exoplanetary field, the same or larger amount of time available to other teams  and 
the necessity to acquire large sets of data to detect the  planetary signals in the radial velocity
time series.
As a consequence of this, most of the published papers are pertinent to the characterization of
planetary systems rather than to the discovery of new planets. However, this should not be considered
a weakness of the GAPS program. The insights into the dynamics of known systems could be more
rewarding than the detection of a planet around another star, especially now that we already know 
more than 2000 planets. The case of the binary system XO-2 illustrates very well this point.
\subsection{The two components of the  XO-2 system}
\begin{figure*}[t!]
\resizebox{\hsize}{!}{\includegraphics[clip=true]{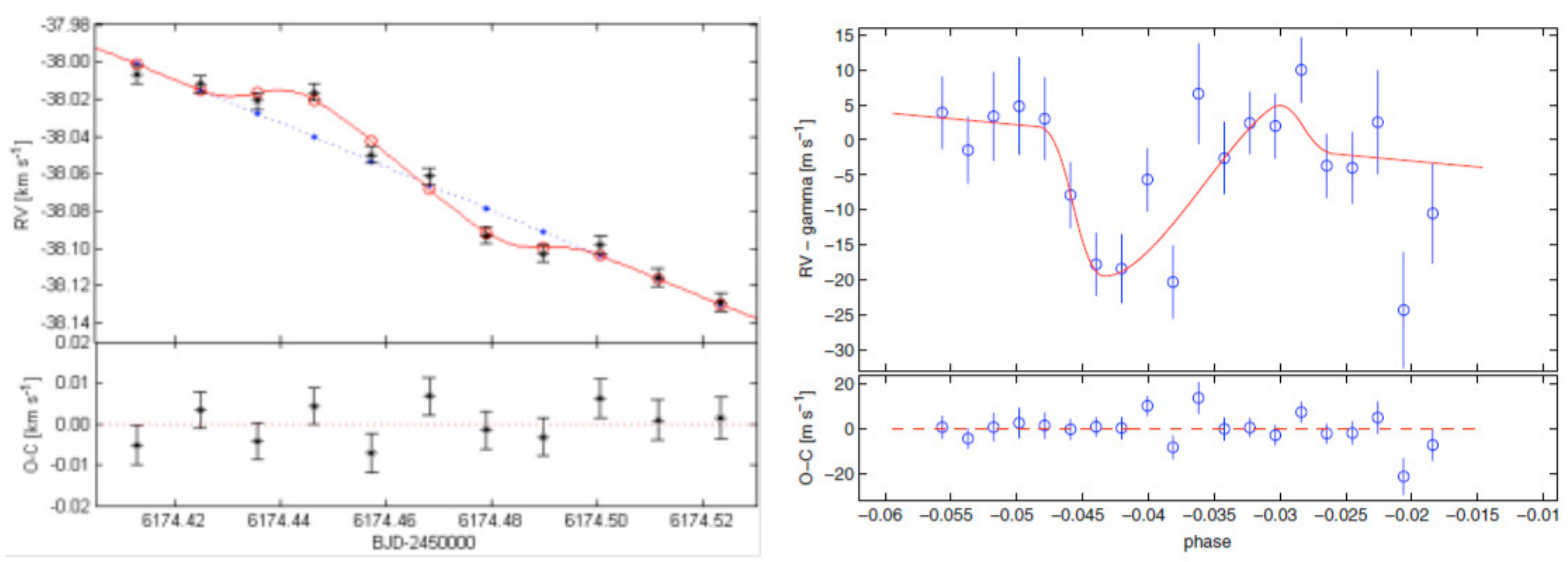}}
\caption{\footnotesize
The RML effects observed on Qatar-1 (left panel) and on HAT-P-18 observed with
HARPS-N by the GAPS team. 
The opposite senses of the variations indicates that 
Qatar-1-b is a classical aligned system, while HAT-P-18-b
is moving on a retrograde orbit.
The residuals of the best-fit solutions are shown in the bottom panels.
}
\label{rml}
\end{figure*}

The formation of planets in binary systems is a challenging and interesting
scientific case. Several planets orbiting one of the binary components are known.
Very wide binaries are the most suitable targets to
search for planetary systems around each component. We were rewarded  
particular attention to the XO-2 system, having  projected separations of 31\arcsec\, and  4000~AU.
The northern component XO-2N was known to host a very close 0.5~$M_J$ transiting 
planet (Fig.~\ref{xo2ns}, left panel), while nothing was known about the southern one, XO-2S. Sixty-three spectra
taken between April 2013 and May 2014 allowed us to discover two planets orbiting
around XO-2S, and a hint of a third, long-period body \citep{desidera}. Their masses
are of the size of Jupiter and Saturn, in orbits wider than that of the planet around XO-2N
(Fig.~\ref{xo2ns}, right panel). 
The Rossiter-McLaughlin effect (see Sect.~\ref{Rml}) was observed for 
the planet transiting in front of XO-2N with 
HARPS-N, and a detailed spectroscopic 
investigation pointed out the differences between the two planetary systems and
an abundance of iron greater in the star XO-2N than in XO-2S 
\citep{xo2dama}.
The latter
fact was corroborated by a further complete abundance analysis that 
revealed the significance of the XO-2N abundance difference relative to XO-2S at 
the $2\sigma$ level for almost all
elements \citep{biazzo}. This result could be interpreted in two ways: 
{\it i)} XO-2N is richer in metals than XO-2S since it ingested close  planets; 
{\it ii)} XO-2S has a depleted composition since  heavy elements are segregated
into the distant  planets   \citep{biazzo}. 
The multitask approach adopted by the GAPS collaboration was very effective in 
investigating this binary system, whose relevance is confirmed by 
the growing interest manifested by other groups
\citep{ramirez, teske}.  
\subsection{The Rossiter-McLaughlin effect}\label{Rml}

\citet{ross} and \citet{mcl} discovered an apparent rotational effect
when observing the eclipses of the $\beta$~Lyr and $\beta$~Per stellar systems, respectively.
Nowadays this effect is thought to be a probe of exoplanet dynamical histories. It
is measured using in-transit spectroscopic observations,
revealing a deviation of the measured radial velocities from the Keplerian orbital motion. 
This apparent deviation is due to the planet occulting part of the rotating stellar surface, thus
introducing an asymmetry in the profiles of the stellar absorption lines. 
The RM waveform
allows us to determine the sky projected spin-orbit alignment angle $\lambda$ between
the rotation axis of the host star and the normal to the planetary
orbital plane. 
The RM effect was recorded in the Sun due to the transit of Venus on June 6, 2012 \citep{molaro}. 
The integrated sunlight as reflected by the Moon at night time was used to follow the transit 
by means of HARPS spectrograph mounted at the 3.6-m ESO telescope.  
The partial eclipse of the solar disc in correspondence of the passage of Venus 
in front of the receding hemisphere produced a modulation in the radial velocity with an 
amplitude of 1~m\,s$^{-1}$, in agreement with the theoretical model. 
This detection anticipates the study of transits of Earth-size bodies in 
solar-type stars by means of a high-resolution spectrograph attached to a 40-m class telescope. 

In the current exoplanetary researches, the alignment angle is thought to provide a window on
the dynamical evolution of exoplanets. Indeed, the
measurements of stellar obliquity have revealed up to now
$\sim$100 systems 
showing a wide range of configurations. In addition to classical
aligned systems \citep[e.g., HD 189733;][]{winn} and
orbits with moderate tilts \citep[e.g., XO-3;][]{hirano}, some
more intriguing cases were observed, as  orbital planes that are perpendicular
to the star's rotational direction \citep[e.g., WASP-7;][]{alb} and
retrograde systems in which the planet orbits in the opposite direction with respect to the
star's rotation \citep[e.g., WASP-17;][]{tri}.
The Kozai-Lidov effect is one of the mechanisms claimed to explain such a 
broad variety of spin-orbit obliquities:
the pertubation by an outer off-plane massive body can induce
periodic oscillations in both the eccentricity and inclination of the inner
planetary orbit. Inward migration of the inner planet then follows, with tidal friction
driving the planet as it approaches its host, causing the orbit to
shrink and circularize \citep[][ and references therein]{fab}.

The GAPS program includes the observation of several transiting planets
showing the RM effect.
\citet{covino} reported about the case of Qatar-1~b, a hot Jupiter
orbiting a metal-rich K-dwarf star on a nearly zero orbital obliquity
(Fig.~\ref{rml}, left panel). The new, very
precise HARPS-N observations allowed us to determine that the planet is significantly
more massive than previously reported. A good sky-projected
spin-orbit alignment was observed for both HAT-P-36-b and WASP-11-b$\equiv$HAT-P-10-b, too
\citep{mancini}.
In the case of HAT-P-36, we could benefit from a coordinated intensive photometric monitoring  
to determine the rotational period of the star and hence 
not only the projected angle, but precisely the true spin-orbit obliquity.
On the other hand
the case of HAT-P-18-b \citep{esposito} was a very peculiar one, since the hot Saturn-mass planet
orbits the late-K dwarf star with a retrograde motion (Fig.~\ref{rml}, right panel).

\subsection{Asteroseismology of the host stars}
The use of asteroseismology in the exoplanetary science is a recurrent subject in
the participation of stellar astronomers to the National Conferences on Planetary Science
since the launch of CoRoT \citep[e.g., ][]{circeo}.
The interplay between asteroseismology and planetary
system  evolution is one of the most innovative aspects of the ESA M4-mission Plato~2.0 \citep{rauer}.
This goal is pursued by the GAPS project, but the reduction of the observing time 
 with respect to the original request did not allowed us to perform the detailed planned
study. However, the bright target $\tau$~Boo~A, hosting a close hot Jupiter, was monitored
as a pilot study.   

Despite the limited time spent on the target, the signature of solar-like oscillations
was detected in the radial-velocities time series obtained with HARPS-N on $\tau$~Boo~A.
We could estimate the frequency of maximum power of the oscillations $\nu_{\rm max}$ 
and the large separation $\Delta\nu$ \citep{borsa}. Both our determinations
agree well with the theoretical predictions. Therefore, we could constrain
the value of the stellar mass to  a 4\% uncertainty (1.38$\pm$0.05 $M_{\odot}$) and derive
the young age of the system (0.9$\pm$0.5~Gyr).

\subsection{Star-planet interaction}
$\tau$~Boo~A has been a very suitable target for the study of the star-planet
interaction. The presence of high-latitude plage was probably detected during
HARPS-N observations, while the correlation betwen the chromospheric activity
and the orbital phase remains unclear \citep{borsa}. We were more successful with
the coordinated HARPS-N and XMM observations of HD~17156 and its planet in a
highly eccentric orbit. 
A significant (6$\sigma$) X-ray brigthening occurred near a periastron passage of the
hot Jupiter, accompanied by an increase of the Ca~II~H\&K chromospheric index \citep{maggio}.

In a parallel study to GAPS activities, \citet{jesus} calibrated some empirical relationships
with the aim  to determine accurate stellar parameters
for M0-M4.5 dwarfs by means of the same spectra used to measure stellar radial velocities.
\subsection{Solving unclear cases}
One of the first results of our HARPS-N survey was the 
confutation of the existence of a multiple giant-planet system around the very metal-poor
star HIP11952 \citep{hip}, which constituted a severe challenge for the current
planet-formation models. Later, we demonstrated that the radial-velocity amplitude
of TrES-4 is smaller than the previously announced one.
Combined with the large radius inferred from the GAPS analysis, TrES-4b becomes the 
second lowest-density transiting hot-Jupiter known \citep{soz}.

\subsection{The discovery of new planets}

We already reported the discovery of two planets around XO-2S and the consequent
great attention of the community around this binary system. We note that the analysis of 
the available radial-velocity data also pointed out possible new long-period planets 
around both XO-2 components \citep{xo2dama}. A new, long-period planet was recently discovered around
Kelt-6 \citep{kelt}.
The decision to be conservative
in announcing new planets, in order to report robust discoveries, is now paying since
we have strong candidates for targets in open clusters (Malavolta et al.,
in prep.), M-dwarfs (Affer et al., in prep.) and giant stars (Micela et al., in prep.). 

\section{Conclusions}

The scientific success of the observational program is certified by the series
of papers on refereed journals (10 accepted up to now, 1 submitted, others in
preparation). This compares favourably with what had been made
within the first three years of operation by the HARPS@ESO GTO programme 
(8 published papers).  But more important of this, 
the GAPS project has been a success in making the Italian team able to reach the 
critical mass required to get visibility in the international context. 
We also put large efforts in the Outreach and Communication activities, with regular
press releases. 
Many of the GAPS papers have a young researcher as first author and many PhD 
students are being forming around the GAPS project. The GAPS 
visibility and  school are of strategic importance when considering the preparation of 
future space missions devoted to the exoplanetary science as CHEOPS and Plato 2.0.

\begin{acknowledgements}
The whole GAPS community acknowledges support from the ``Progetti Premiali" funding scheme of the Italian
Ministry of Education, University, and Research.
\end{acknowledgements}

\bibliographystyle{aa}

\end{document}